\documentclass[english,aps,preprint,superscriptaddress]{revtex4-1}
\usepackage{lmodern}
\usepackage{lmodern}
\usepackage[T1]{fontenc}
\usepackage[utf8]{inputenc}
\usepackage{amsmath}
\usepackage{amsthm}
\usepackage{amssymb}
\usepackage{graphicx}
\usepackage{esint}

\makeatletter

\usepackage{amsthm}\usepackage{latexsym}\usepackage{bm}\usepackage{amsfonts}

\usepackage{babel}
\usepackage{hyperref}
\usepackage{enumitem}

\makeatother

\usepackage{babel}
\begin{document}
\title{Kelvin-Helmholtz instability is the result of parity-time symmetry
breaking}
\author{Hong Qin}
\email{hongqin@princeton.edu}

\affiliation{Plasma Physics Laboratory, Princeton University, Princeton, NJ 08543,
U.S.A}
\affiliation{School of Physical Sciences, University of Science and Technology
of China, Hefei, 230026, China}
\author{Ruili Zhang}
\affiliation{School of Science, Beijing Jiaotong University, Beijing 100044, China}
\author{Alexander S. Glasser}
\affiliation{Plasma Physics Laboratory, Princeton University, Princeton, NJ 08543,
U.S.A}
\author{Jianyuan Xiao}
\affiliation{School of Physical Sciences, University of Science and Technology
of China, Hefei, 230026, China}
\begin{abstract}
Parity-Time (PT)-symmetry is being actively investigated as a fundamental
property of observables in quantum physics. We show that the governing
equations of the classical two-fluid interaction and the incompressible
fluid system are PT-symmetric, and the well-known Kelvin-Helmholtz
instability is the result of spontaneous PT-symmetry breaking. It
is expected that all classical conservative systems governed by Newton's
law admit PT-symmetry, and the spontaneous breaking thereof is a generic
mechanism for classical instabilities. Discovering the PT-symmetry
of systems in fluid dynamics and plasma physics and identifying the
PT-symmetry breaking responsible for instabilities enable new techniques
to classical physics and enrich the physics of PT-symmetry.
\end{abstract}

\maketitle
Parity-Time (PT)-symmetry \citep{bender1998real,bender2002complex,bender2007making,Bender2010,Lee69,mostafazadeh2002pseudo,mostafazadeh2002pseudoII,mostafazadeh2002pseudoIII,Zhang18}
is an important topic in quantum mechanics. It pertains to one of
the fundamental questions for quantum theory, i.e., what are observables?
The accepted consensus is that observables are Hermitian operators
in Hilbert spaces, and, as a consequence, their eigenvalues are real.
However, observables need not to be Hermitian to have real eigenvalues.
Bender and collaborators \citep{bender1998real,bender2002complex,bender2007making,Bender2010}
proposed the concept of PT-symmetric operators to relax the Hermitian
requirement. Specifically, consider Schrödinger's equation specified
by a Hamiltonian $H$,
\begin{equation}
\dot{\psi}=-iH\psi=A\psi\thinspace,\label{eq:ls}
\end{equation}
where $iA$ is another notation for $H$. 

The Hamiltonian $H$ is called PT-symmetric if 
\begin{equation}
PTH-HPT=0\thinspace,\label{eq:PT}
\end{equation}
where $P$ is a linear operator satisfying $P^{2}=I$ and $T$ is
the complex conjugate operator \citep{bender2007making}. In finite
dimensions, which will be the focus of the present study, $H$, $A$
and $P$ can be represented by matrices, and the PT-symmetric condition
\eqref{eq:PT} is equivalent to 
\begin{equation}
P\bar{H}-HP=0\textrm{ \thinspace\thinspace or\thinspace\thinspace\ }PA+\bar{A}P=0\thinspace,\label{eq:PT1}
\end{equation}
Here, $\bar{H}$ and $\bar{A}$ denote the complex conjugates of $H$
and $A$, respectively. The PT-symmetry condition can be understood
as follows. In terms of a $P$-reflected variable $\phi\equiv P\psi$,
Eq.\,\eqref{eq:ls} is 
\begin{equation}
\dot{\phi}=PAP^{-1}\phi\thinspace.\label{eq:y}
\end{equation}
In general, Eq.\,\eqref{eq:ls} is not $P$-symmetric, i.e., $PAP^{-1}\neq A$.
However, we can check the effect of applying an additional $T$-reflection,
i.e., $t\rightarrow-t$ and $i\rightarrow-i$. Then Eq.\,\eqref{eq:y}
becomes 
\begin{equation}
\dot{\bar{\phi}}=-\overline{PAP^{-1}}\bar{\phi}\thinspace.
\end{equation}
If $-\overline{PAP^{-1}}=A$, which is equivalent to the PT-symmetric
condition \eqref{eq:PT} or \eqref{eq:PT1}, then Eq.\,\eqref{eq:y}
in terms of $\bar{\phi}$ is identical to Eq.\,\eqref{eq:ls} in
terms of $\psi.$ Thus, PT-symmetry is an invariant property of the
system under the reflections of both parity and time. 

The spectrum of a PT-symmetric Hamiltonian $H$ has the following
properties \citep{bender1998real,bender2002complex,bender2007making,Bender2010}:
\begin{enumerate}
\item The spectrum is symmetric with respect to the real axis. In another
word, $\lambda=a+ib$ is an eigenvalue of $H$ if and only if $\lambda=a-ib$
is. 
\item There are boundaries in the parameter space that separate regions
with unbroken PT-symmetry where all eigenvalues of $H$ are real from
regions with broken PT-symmetry where $H$ has at least one pair of
complex conjugate eigenvalues. The boundaries are also known as exceptional
points \citep{heiss2004exceptional,heiss2012physics}.
\item In regions with unbroken PT-symmetry, any eigenvector $\xi$ of $H$
is also an eigenvector of the $PT$ operator, i.e., $(PT)\xi=\alpha\xi,$
which implies that the PT-symmetry of $H$ is preserved by the eigenvector
or eigenmode. In regions with broken PT-symmetry, the eigenvectors
corresponding to the pair of complex conjugate eigenvalues do not
preserve the PT-symmetry, i.e., they are not eigenvectors of the $PT$
operator. 
\end{enumerate}
Note that in the terminology adopted by the community, PT-symmetry
breaking does not mean that Eq.\,(\ref{eq:PT}) is violated. It only
means that one of the eigenvector does not preserve the PT-symmetry.
This type of symmetry breaking is known as spontaneous symmetry breaking
in quantum theory.

While Properties (I) and (III) of the spectrum are straightforward
to prove from Eqs.\,(\ref{eq:PT}) or (\ref{eq:PT1}) \citep{bender1998real,bender2002complex,bender2007making,Bender2010},
how PT-symmetry breaking happens is still an active research topic
\citep{heiss2004exceptional,heiss2012physics,Zhang18}. For example,
it was recently proved that a PT-symmetric Hamiltonian in finite dimensions
is necessarily pseudo-Hermitian regardless whether it is diagonalizable
or not \citep{Zhang18}, and a necessary and sufficient condition
for PT-symmetry breaking is the resonance between a positive-action
mode and a negative-action mode \citep{Zhang16GH,Zhang1711,Zhang18,kirillov2017singular,Krein50,Gelfand55,Yakubovich75}. 

Since it inception, the concept of PT-symmetry has been quickly extended
to classical physics \citep{El-Ganainy2007,Guo2009,klaiman2008visualization,Chong2011,peng2014parity,hodaei2017enhanced,brandstetter2014reversing,schindler2011experimental,bender2013twofold,Kirillov2013,bender2014systems,Bender2016,Tsoy2017}.
Classical systems with PT-symmetry have been discovered, and the systems
are stable when the PT-symmetry is unbroken and unstable when the
PT-symmetry is broken. Identifying PT-symmetry breaking as a fundamental
mechanism for classical instabilities brings new perspectives and
methods to classical physics. 

Many of the classical systems with PT-symmetry studied so far \citep{schindler2011experimental,bender2013twofold,Kirillov2013,bender2014systems,Bender2016,Tsoy2017},
especially those in mechanics and electrical circuits, are real canonical
Hamiltonian systems in the form of 
\begin{align}
\dot{\boldsymbol{z}} & =JE\boldsymbol{z},\\
J & =\left(\begin{array}{cc}
0 & I_{n}\\
-I_{n} & 0
\end{array}\right)\,,
\end{align}
where $\boldsymbol{z}\in\mathbb{R}^{2n}$, $J$ is the standard $2n\times2n$
symplectic matrix, and $E$ is a $2n\times2n$ real symmetric matrix.
The matrix $JE$ plays the role of $A$ in Schrödinger's equation
(\ref{eq:ls}). For such a real Hamiltonian system, it well-known
that the spectrum of $JE$ is symmetric with respect to the imaginary
axis, or equivalently, the spectrum of $iJE$ is symmetric with respect
to the real axis. The boundaries between stable and unstable regions
in the parameter are locations for the Hamilton-Hopf bifurcation \citep{Qin15-056702,Qin2018}.
Thus, real canonical Hamiltonian systems are naturally endowed with
Properties (I) and (II) listed above, whereas Property (III) is uniquely
associated with PT-symmetry. On the other hand, the PT-symmetry found
in optical and microwave devices is in general for complex dynamical
systems \citep{El-Ganainy2007,Guo2009,klaiman2008visualization,Chong2011,peng2014parity,hodaei2017enhanced,brandstetter2014reversing,Bender2016a}.
For these systems, PT-symmetry is often identified as a balanced loss-gain
mechanism. However, it is probably more appropriate and general to
consider PT-symmetry as a fundamental symmetry property that physical
systems possess, other than a specific mechanism for balancing energy. 

In this paper, we show that the governing equations for the classical
Kelvin-Helmholtz (KH) instability in fluid dynamics is a complex system
with a PT-symmetry, which is associated with neither the real canonical
Hamiltonian structure nor the balanced loss-gain mechanism. The KH
instability is probably the most famous instability in classical systems.
It is a representative of classical instabilities in fluid dynamics
and plasma physics. The discovery of PT-symmetry in this system is
therefore valuable in terms of identifying a new class of problems
for the physics of PT-symmetry. 

Specifically, we show that the fluid equation system governing the
KH instability has the spectrum Properties (I) and (II), and theory
are the consequences of the PT-symmetry admitted by the system. More
fundamentally, the new revelation enabled by the discovery of PT-symmetry
is Property (III), i.e., the KH instability occurs when and only when
the PT-symmetry is spontaneously broken. Establishing the connection
between PT-symmetry and the spectrum properties for classical systems
in fluid dynamics and plasma physics is fully consistent with the
basic philosophy of PT-symmetry in quantum systems, i.e., the spectrum
properties of an observable should be result of the physical requirement
of PT-symmetry, instead of the Hermiticity. As a matter of fact, it
is rare for the Hamiltonians of classical systems to be Hermitian.
Our goal here is to demonstrate, using the example of the KH instability,
that classical conservative systems governed by Newton's law are PT-symmetric,
since the PT-symmetry is a manifestation of the classical reversibility.
And as a consequence, classical instabilities of conservative systems
are the result of spontaneous PT-symmetry breaking.

\begin{figure}[ptb]
\begin{centering}
\includegraphics[width=3in]{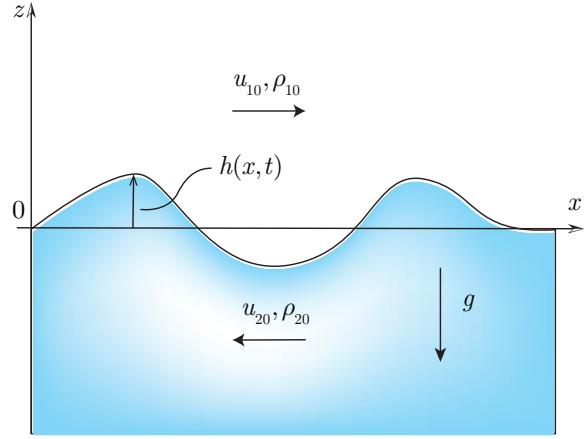} 
\par\end{centering}
\caption{Kelvin-Helmholtz instability is the result of PT-symmetry breaking.}

\label{KH} 
\end{figure}

To facilitate the discussion, we present here a brief but complete
derivation of the classical KH instability. Consider a 2D incompressible
fluid system in a gravitational field described by the following set
of equations, 
\begin{align}
\frac{\partial u}{\partial x}+\frac{\partial w}{\partial z} & =0\,,\label{eq:f1}\\
\frac{\partial u}{\partial t}+u\frac{\partial u}{\partial x}+w\frac{\partial u}{\partial z} & =-\frac{1}{\rho}\frac{\partial p}{\partial x}\,,\\
\frac{\partial w}{\partial t}+u\frac{\partial w}{\partial x}+w\frac{\partial w}{\partial z} & =-\frac{1}{\rho}\frac{\partial p}{\partial z}-g\,,\\
\frac{\partial\rho}{\partial t}+u\frac{\partial\rho}{\partial x}+w\frac{\partial\rho}{\partial z} & =0\,.\label{eq:f4}
\end{align}
Here, $(u,w)$ is the velocity field in the $(x,z)$ plane, $\rho$
is density, $p$ is the pressure, and the gravitational field $g$
is in the direction of $-z$. The unperturbed equilibrium consists
of two layers of fluid $(i=1,2)$ separated at $z=0$ with 
\begin{align}
u_{i0} & =const.\,,\thinspace w_{i0}=0\,,\thinspace\rho_{i0}=const.\,,\\
p_{i0} & =p_{0}-\rho_{i0}gz\,,
\end{align}
In general, $u_{10}\neq u_{20}$ and $\rho_{10}\neq\rho_{20}.$ Consider
a perturbation of the system of the form 
\begin{align}
u_{i} & =u_{i0}+\delta u_{i}\,,\thinspace w_{i}=\delta w_{i}\,,\,\rho_{i}=\rho_{i0}\,,\\
p_{i} & =p_{0}-\rho_{i0}gz+\delta p_{i}\,,
\end{align}
where $(i=1,2)$. Let's denote the perturbed boundary between the
two layers of fluid by $h(x,t).$ The equilibrium and perturbation
of the system are illustrated in Fig.~\ref{KH}. 

Since the equilibrium is homogeneous in the $x$-direction, different
Fourier components in the $x$-direction are decoupled, and we can
investigate each Fourier component separately. Assuming $(\delta u_{i},\delta w_{i},\delta p_{i})\sim\exp(ikx)$
for a wavelength $k\in\mathbb{R},$ and linearizing the fluid system
(\ref{eq:f1})-(\ref{eq:f4}), we obtain
\begin{align}
ik\delta u_{i}+\frac{\partial\delta w_{i}}{\partial z} & =0\,,\label{eq:d1}\\
\frac{\partial\delta u_{i}}{\partial t}+u_{i0}ik\delta u_{i} & =-\frac{ik}{\rho_{i0}}\delta p_{i}\,,\label{eq:d2}\\
\frac{\partial\delta w_{i}}{\partial t}+u_{i0}ik\delta w_{i} & =-\frac{1}{\rho_{i0}}\frac{\partial\delta p_{i}}{\partial z}\,.\label{eq:d3}
\end{align}
In Eqs.\,(\ref{eq:d1})-(\ref{eq:d3}), $(\delta u_{i},\delta w_{i},\delta p_{i})$
represent the complex Fourier components of the perturbed fields at
the wavelength $k$ in the $x$-direction. To simplify the notation,
we have overloaded the symbols $(\delta u_{i},\delta w_{i},\delta p_{i})$
to represent both the real perturbed fields on $(x,z)$, as well as
their complex amplitudes on the spectrum space. Simple manipulation
of Eqs.\,(\ref{eq:d1})-(\ref{eq:d3}) shows that
\[
\frac{\partial^{2}}{\partial z^{2}}\left(\frac{\partial\delta w_{1}}{\partial t}+iku_{10}\delta w_{1}\right)=\frac{\partial^{2}}{\partial z^{2}}\left(\frac{\partial\delta w_{2}}{\partial t}+iku_{20}\delta w_{2}\right)\,,
\]
which implies that 
\begin{align}
(\delta w_{1},\delta p_{1}) & \sim\exp(ikx-\left|k\right|z)\,,\label{eq:xz1}\\
(\delta w_{2},\delta p_{2}) & \sim\exp(ikx+\left|k\right|z)\,,\label{eq:xz2}
\end{align}
where the boundary condition $\lim_{z\rightarrow\infty}\delta w_{1}=\lim_{z\rightarrow-\infty}\delta w_{2}=0$
has been incorporated. Note that the wavelength $k$ is allowed to
be negative. With the $x$- and $z$-dependency specified by Eqs.\,(\ref{eq:xz1})
and (\ref{eq:xz2}), Eqs.\,(\ref{eq:d1})-(\ref{eq:d3}) for the
two layers of fluid $(i=1,2)$ reduce to, respectively, 
\begin{align}
ik\delta u_{1}-\left|k\right|\delta w_{1} & =0\,,\label{eq:e1}\\
\frac{\partial\delta u_{1}}{\partial t}+iku_{10}\delta u_{1} & =-\frac{ik}{\rho_{10}}\delta p_{1}\,,\label{eq:e2}\\
\frac{\partial\delta w_{1}}{\partial t}+iku_{10}\delta w_{1} & =\frac{\left|k\right|}{\rho_{10}}\delta p_{1}\,,\label{eq:e3}
\end{align}
 and 
\begin{align}
ik\delta u_{2}+\left|k\right|\delta w_{2} & =0\,,\label{eq:e1-1}\\
\frac{\partial\delta u_{2}}{\partial t}+iku_{20}\delta u_{2} & =-\frac{ik}{\rho_{20}}\delta p_{2}\,,\label{eq:e2-1}\\
\frac{\partial\delta w_{2}}{\partial t}+iku_{20}\delta w_{2} & =-\frac{\left|k\right|}{\rho_{20}}\delta p_{2}\,.\label{eq:e3-1}
\end{align}
Now we invoke the boundary conditions at the interface, 

\begin{align}
\frac{d}{dt}\left(h(x,t)-z\right) & =u_{i0}\frac{\partial h}{\partial x}+\frac{\partial h}{\partial t}-\frac{dz}{dt}=0,\,\,(i=1,2)\,,\\
p_{1}\mid_{z=h} & =p_{2}\mid_{z=h}\,,
\end{align}
which, in terms of the perturbed fields, are
\begin{align}
u_{i0}ikh+\frac{\partial h}{\partial t}-\delta w_{i} & =0\,,\,\,(i=1,2)\,,\label{eq:b1}\\
-\rho_{10}g+\delta p_{1} & =-\rho_{20}g+\delta p_{2}\,,\label{eq:b2}
\end{align}
Define $\phi_{1}\equiv-\delta w_{1}/\left|k\right|$ and $\phi_{2}\equiv\delta w_{2}/\left|k\right|$.
Equations (\ref{eq:e1})-(\ref{eq:e3-1}), (\ref{eq:b1}) and (\ref{eq:b2})
can be combined to give the governing ODEs for $\phi_{1}$, $\phi_{2},$
and $h,$
\begin{align}
\phi_{1} & =-\frac{1}{\left|k\right|}\left(\frac{\partial}{\partial t}+iku_{10}\right)h\,,\label{eq:phi1}\\
\phi{}_{2} & =\frac{1}{\left|k\right|}\left(\frac{\partial}{\partial t}+iku_{20}\right)h\,,\label{eq:phi2}\\
\rho_{10}\left(\frac{\partial\phi_{1}}{\partial t}+iku_{10}\phi_{1}+gh\right) & =-\rho_{20}\left(\frac{\partial\phi_{2}}{\partial t}+iku_{20}\phi_{2}+gh\right)\,.\label{eq:rhi1}
\end{align}
Assuming $(\phi_{1},\phi_{2},h)\sim\exp(-i\omega t)$, we obtain from
Eqs.\,(\ref{eq:phi1})-(\ref{eq:rhi1}) the dispersion relation of
the dynamics, 
\begin{equation}
\omega_{\pm}=k\frac{\rho_{10}u_{10}+\rho_{20}u_{20}}{\rho_{10}+\rho_{20}}\pm\sqrt{\left|k\right|g\frac{\rho_{20}-\rho_{10}}{\rho_{20}+\rho_{10}}-\frac{k^{2}\rho_{10}\rho_{20}(u_{10}-u_{20})^{2}}{(\rho_{10}+\rho_{20})^{2}}}\,.\label{eq:DR}
\end{equation}
For a given $k\in\mathbb{R},$ when $g(\rho_{20}^{2}-\rho_{10}^{2})>\left|k\right|\rho_{10}\rho_{20}(u_{10}-u_{20})^{2}$,
the eigen-frequencies $\omega_{\pm}$ of the two-fluid interaction
are real and the dynamics is stable. When $g(\rho_{20}^{2}-\rho_{10}^{2})<\left|k\right|\rho_{10}\rho_{20}(u_{10}-u_{20})^{2}$,
the eigen-frequencies $\omega_{\pm}$ are a pair of complex conjugate
numbers and the dynamics is unstable. This is the well-known KH instability.
Since the gravity is considered, it includes the Rayleigh-Taylor instability
as a special case. For this reason, it can also be called Kelvin-Helmholtz-Rayleigh-Taylor
instability. The boundary between stability and instability is at
$g(\rho_{20}^{2}-\rho_{10}^{2})=\left|k\right|\rho_{10}\rho_{20}(u_{10}-u_{20})^{2}$.
Obviously, these spectrum properties of the KH instability are exactly
the Properties (I) and (II) for a PT-symmetric system. This observation
naturally begs the question: is the system governing the KH instability
PT-symmetric?

The answer is affirmative. This discovery reveals, according to Property
(III), that the classical KH instability is the result of PT-symmetry
breaking. Now we prove these facts explicitly and explain the interesting
physics underpinning the mathematics. 

Let's start from the governing ODEs (\ref{eq:phi1})-(\ref{eq:rhi1}),
which do not assume the form Eq.\,(\ref{eq:ls}), the standard form
for dynamic systems. A few lines of algebra show that Eqs.\,(\ref{eq:phi1})-(\ref{eq:rhi1})
actually describe a system with two independent complex variables.
Therefore, we use Eq.\,(\ref{eq:phi2}) to eliminate $\phi_{2}$
in favor of $\phi_{1}$ and $h,$ and obtain
\begin{align}
\left(\begin{array}{c}
\dot{\phi}_{1}\\
\dot{h}
\end{array}\right) & =A\left(\begin{array}{c}
\phi_{1}\\
h
\end{array}\right)\,,\label{eq:phidot}\\
A & =\left(\begin{array}{cc}
\dfrac{ik\left(-u_{10}\rho_{10}-2u_{20}\rho_{20}+u_{10}\rho_{20}\right)}{\rho_{10}+\rho_{20}} & \dfrac{-\left|k\right|(u_{10}-u_{20})^{2}\rho_{20}+g(\rho_{20}-\rho_{10})}{\rho_{10}+\rho_{20}}\\
-\left|k\right| & -iku_{10}
\end{array}\right)\,.\label{eq:AKH}
\end{align}
It can be verified that Eq.\,(\ref{eq:phidot}) is invariant under
the following PT-transformation,

\begin{equation}
\left(\phi_{1},h,t,i\right)\rightarrow\left(-\phi_{1},h,-t,-i\right)\,.\label{eq:PT-1}
\end{equation}
In terms of matrix $A,$ this PT-symmetry is $PA+\bar{A}P=0$, i.e.,
Eq.\,(\ref{eq:PT1}), for
\begin{equation}
P=\left(\begin{array}{cc}
-1 & 0\\
0 & 1
\end{array}\right)\,.
\end{equation}
We have thus proved that governing equations of the KH instability
is indeed PT-symmetric. This explains the spectrum Properties (I)
and (II) of the system, and more importantly, stipulates that the
KH instability occurs when and only when the PT-symmetry is broken.

To explicitly verify the PT-symmetry breaking as the mechanism for
instability, we look at the eigenvectors $v_{\pm}$ of $H=iA,$
\begin{align}
v_{\pm}= & \left(\begin{array}{c}
\pm i\bar{\phi}\\
1
\end{array}\right)\,,\\
\bar{\phi}= & \frac{k(u_{20}-u_{10})\rho_{20}}{\left|k\right|(\rho_{10}+\rho_{20})}+\frac{1}{\left|k\right|}\sqrt{\left|k\right|g\frac{\rho_{20}-\rho_{10}}{\rho_{20}+\rho_{10}}-\frac{k^{2}\rho_{10}\rho_{20}(u_{10}-u_{20})^{2}}{(\rho_{10}+\rho_{20})^{2}}}\,.
\end{align}
When the system is stable, i.e., $g(\rho_{20}^{2}-\rho_{10}^{2})>\left|k\right|\rho_{10}\rho_{20}(u_{10}-u_{20})^{2}$,
both eigenvectors $v_{\pm}$ preserve the PT-symmetry, which means
that they are invariant under the PT-transformation (\ref{eq:PT-1}).
When the KH instability occurs, both eigenvectors are not invariant
under the PT-transformation (\ref{eq:PT-1}), and the PT-symmetry
is broken. As a matter of fact, the PT-transformation (\ref{eq:PT-1})
maps $v_{+}$ into $v_{-}$, and vice versa. 

In terms of physics, the PT-symmetry of Eq.\,(\ref{eq:phidot}) can
be traced all the way back to the fluid equations (\ref{eq:f1})-(\ref{eq:f4}).
We observe that the system is symmetric with respect to the following
PT-transformation,
\begin{equation}
(t,u,w,p,\rho)\rightarrow(-t,-u,-w,p,\rho)\,.\label{eq:PT-2}
\end{equation}
This PT-symmetry is a discrete symmetry originates from the reversibility
of the fluid system, which is expected for conservative classical
systems governed by Newton's second law in fluid dynamics and plasma
physics. When the solutions of the system preserve this PT-symmetry,
i.e., the PT-symmetry is unbroken, the dynamics is stable. On the
other hand, when a solution is not invariant under the PT-transformation
(\ref{eq:PT-2}), the PT-symmetry is (spontaneously) broken and the
system is unstable. This is the mechanism of the KH instability. Apparently,
this instability mechanism applies to all instabilities that can be
described by the fluid equations (\ref{eq:f1})-(\ref{eq:f4}). The
KH instability is just the one of them, admittedly the most famous
one. 

It should be evident that parity transformation will have a different
form for different choices of dynamic variables. For example, we can
define a pair of new variables as
\begin{equation}
\left(\begin{array}{c}
\psi_{1}\\
\psi_{2}
\end{array}\right)=Q\left(\begin{array}{c}
\phi_{1}\\
h
\end{array}\right)\,,\,\,\,Q\equiv\left(\begin{array}{cc}
-1 & u_{10}\\
1 & u_{10}
\end{array}\right)\,.
\end{equation}
 In terms of $\psi_{1}$ and $\psi_{2}$, Eq.\,(\ref{eq:phidot})
is 
\begin{equation}
\left(\begin{array}{c}
\dot{\psi}_{1}\\
\dot{\psi}_{2}
\end{array}\right)=QAQ^{-1}\left(\begin{array}{c}
\psi_{1}\\
\psi_{2}
\end{array}\right)\,.\label{eq:psi}
\end{equation}
Equation (\ref{eq:psi}) is symmetric with respect to the following
PT-transformation 
\[
\left(\psi_{1},\psi_{2},t,i\right)\rightarrow\left(\psi_{2},\psi_{1},-t,-i\right)\,,
\]
which is probably more familiar than the one given by Eq.\,(\ref{eq:PT-1}).
But the difference is merely cosmetic. 

Let's finish our discussion with a few footnotes. First, we expect
that all classical conservative systems governed by Newton's law are
PT-symmetric, since the PT-symmetry is a manifestation of the classical
reversibility. As a consequence, we further expected all classical
instabilities of conservative systems are the result of spontaneous
PT-symmetry breaking. Secondly, the PT-symmetry can also be broken
explicitly by the governing equations, as oppose to spontaneously
by the solutions. This happens, for instance, when the system is dissipative
due to mechanical frictions or radiation reaction force on charged
particles. Interestingly, explicit PT-symmetry breaking due to dissipation
can also lead to instabilities. Finally, specific classical systems
may admit special PT-symmetries not necessarily associated with classical
reversibility of conservative systems. For example, in stellarators,
a family of non-axisymmetric magnetic confinement devices, PT-symmetry
has been identified by Dewar and collaborators and its effects have
been studied \citep{Dewar1992,Dewar1998} . 

Relevant results in these aspects will be reported in other publications. 
\begin{acknowledgments}
This research was supported by the U.S. Department of Energy (DE-AC02-09CH11466),
the National Natural Science Foundation of China (NSFC-11505186),
and China Postdoctoral Science Foundation (2017LH002).
\end{acknowledgments}

\bibliographystyle{apsrev4-1}
\bibliography{KH}

\end{document}